\documentclass[10pt,letterpaper]{article}         %
\usepackage{opex3}                                %
\usepackage{graphicx}
\usepackage{epsfig}
\usepackage{cite}
\usepackage{color}
\usepackage{amsmath,amssymb,subfigure}

\newcommand{\bra}[1]{\langle #1 |}
\newcommand{\ket}[1]{| #1 \rangle}

\begin{document}
\title{
Mach-Zehnder interferometer using frequency-domain beamsplitter
}
\author{
Toshiki~Kobayashi,$^{1,*}$
Daisuke~Yamazaki,$^{1}$
Kenichiro~Matsuki,$^{1}$
Rikizo~Ikuta,$^{1}$
Shigehito~Miki,$^{2,3}$
Taro~Yamashita,$^{2}$
Hirotaka~Terai,$^{2}$
Takashi~Yamamoto,$^{1}$
Masato~Koashi,$^{4}$ and 
Nobuyuki~Imoto$^{1,**}$}

\address{
$^1$Graduate School of Engineering Science, Osaka University,
Toyonaka, Osaka 560-8531, Japan\\
$^2$Advanced ICT Research Institute, 
National Institute of Information and Communications Technology (NICT),
Kobe 651-2492, Japan\\
$^3$Graduate School of Engineering Faculty of Engineering, Kobe University,
Kobe 657-0013, Japan\\
$^4$Photon Science Center, 
The University of Tokyo, Bunkyo-ku, Tokyo 113-8656, Japan
}

\email{$^*$kobayashi-t@qi.mp.es.osaka-u.ac.jp} 
\email{$^{**}$imoto@mp.es.osaka-u.ac.jp} 



\begin{abstract}
We demonstrated the first-order interference between coherent light at 1580~nm and 795~nm by using frequency-domain Mach-Zehnder interferometer~(MZI).
The MZI is implemented by two frequency-domain BSs based on a second-order nonlinear optical effect in a periodically-poled lithium niobate waveguide with a strong pump light.
We achieved the visibility of over 0.99 at 50\% conversion efficiencies of the BSs.
Toward photonic quantum information processing, 
sufficiently small background photon rate is necessary.
From the measurement results with a superconducting single photon detector~(SSPD),
we discuss the feasibility of the frequency-domain MZI in a quantum regime.
Our estimation shows that single photon interference with the visibility above 0.9 is 
feasible with practical settings.
\end{abstract}


\section{Introduction}
Quantum frequency conversion~(QFC)\cite{Kumar1990} enables a color change of light while preserving its quantum state.
 It has been actively studied for bridging gaps among physical systems and devices each of which prefers its own response frequency, such as upconversion single photon detectors\cite{Langrock2005, Rakher2010, Pelc2011} and photonic quantum memories connecting
to a telecom optical fiber\cite{Tanzilli2005, Dudin2010, Ikuta2011, Zaske2012, Ikuta2013, Ikuta2013-2, Albrecht2014, Farrena2016, Ikuta2016}.
The QFC process is based on $\chi^{(2)}$ or $\chi^{(3)}$ nonlinear optical effect with strong pump lights.
It is described by an effective Hamiltonian of a beamsplitter~(BS) acting on two different frequency modes\cite{Kumar1990, McKinstrie2005}.
Therefore QFC process is expected to work as a frequency-domain BS and plays an essential role for the universal quantum operation on the frequency degree of freedom, similarly to the conventional BS acting on spatial modes in linear optical quantum information processing~(QIP)\cite{Knill2001, Koashi2001}. One of the advantages of frequency domain QIP is the easy access to the frequency multiplexed entangled photon sources demonstrated in Refs.\cite{Li2009, Ramelow2009, Olislager2010,Bernhard2013,Jin2016}. Recently, to see the fundamental property of  frequency-domain BS, 
Hong-Ou-Mandel interference between photons with different frequencies\cite{Kobayashi2016}
which used a partial frequency converter based on $\chi^{(2)}$ nonlinearity in a periodically-poled lithium niobate~(PPLN) waveguide\cite{Ikuta2013:3} has been observed.
The first-order interference of a single photon in different frequencies with near 50\% visibility has also been shown by using
$\chi^{(3)}$ nonlinear process in a 100~m fiber in a cryostat\cite{Clemmen2016}.
On the other hand, the first-order interference by using the $\chi^{(2)}$ nonlinear materials is expected to have a higher signal-to-noise ratio leading to a high visibility even in a room temperature.

In this paper,
we constructed the frequency-domain Mach-Zehnder interferometer~(MZI) employing two frequency-domain BSs based on  PPLN waveguides, which is used as the $\chi^{(2)}$ nonlinear material.
The first frequency-domain BS divides the input coherent light at 1580~nm into 795-nm and 1580-nm light. The second frequency-domain BS combines the two lights and show the interference fringe at each output mode (795-nm and 1580-nm). 
When the conversion efficiency of both frequency-domain BSs was $\sim$ 0.5, the observed visibility was over 0.99.
We also evaluated the performance of the interferometer at a single photon level.
From the results, we found that estimated visibilities will be over 0.9 by using commercially available bandpass filters.

\section{Theory of QFC and frequency-domain Mach-Zehnder interferometer}
We briefly review the dynamics of QFC process in a $\chi^{(2)}$ nonlinear optical medium~\cite{Kumar1990, Ikuta2011}. 
We consider two signal modes at upper and lower angular frequencies $\omega_\mathrm{U}$ and $\omega_\mathrm{L}$ satisfying $\omega_\mathrm{U} = \omega_\mathrm{L} + \omega_\mathrm{P}$, where
$\omega_\mathrm{P}$ is the angular frequency of pump light.
With an assumption of undepletion of the strong pump,
an effective Hamiltonian of the three wave mixing process is described by
\begin{equation}
  \label{eq:h}
  \hat{H}=i\hbar \left(\chi^* \hat{a}_\mathrm{L}^\dagger \hat{a}_\mathrm{U}-\chi \hat{a}_\mathrm{L} \hat{a}_\mathrm{U}^\dagger \right),
\end{equation}
where $\hat{a}_\mathrm{U}$ and $\hat{a}_\mathrm{L}$ are annihilation operators of the upper and lower frequency modes,
respectively.
Here $\chi = |\chi| e^{i\phi_\mathrm{P}}$ is proportional to the complex amplitude of the pump light,
where $\phi_\mathrm{P}$ represents the phase of the pump light.
Using the Heisenberg equation from the above Hamiltonian,
annihilation operators $\hat{a}_\mathrm{U,out}$ and $\hat{a}_\mathrm{L,out}$ of upper and lower modes 
output from the nonlinear optical medium are described by
\begin{equation}
  \label{eq:upper}
  \hat{a}_\mathrm{U,out}= \hat{U}^\dagger \hat{a}_\mathrm{U}\hat{U}=
   \cos(|\chi|\tau)\hat{a}_\mathrm{U}-e^{i\phi_\mathrm{P}}\sin(|\chi|\tau)\hat{a}_\mathrm{L}
\end{equation}
and
\begin{equation}
  \label{eq:lower}
  \hat{a}_\mathrm{L,out}= \hat{U}^\dagger \hat{a}_\mathrm{L}\hat{U}=
  e^{-i\phi_\mathrm{P}} \sin(|\chi|\tau)\hat{a}_\mathrm{U}+\cos(|\chi|\tau)\hat{a}_\mathrm{L},
\end{equation}
where $\hat{U}\equiv \exp(-i \hat{H}\tau/\hbar)$ and  $\tau$ is the traveling time of the light through the $\chi^{(2)}$ medium.

Eqs.~(\ref{eq:upper}) and (\ref{eq:lower}) are regarded as the dynamics of a BS with
the transmittance $T = |\cos(|\chi|\tau)|^2$ and the reflectance $R= |\sin(|\chi|\tau)|^2$ on the two modes of $\hat{a}_\mathrm{U}$ and $\hat{a}_\mathrm{L}$.
They can be actively adjusted by changing the pump power. 
\begin{figure*}[t]
 \begin{center}
  \includegraphics[bb=0 0 308 165]{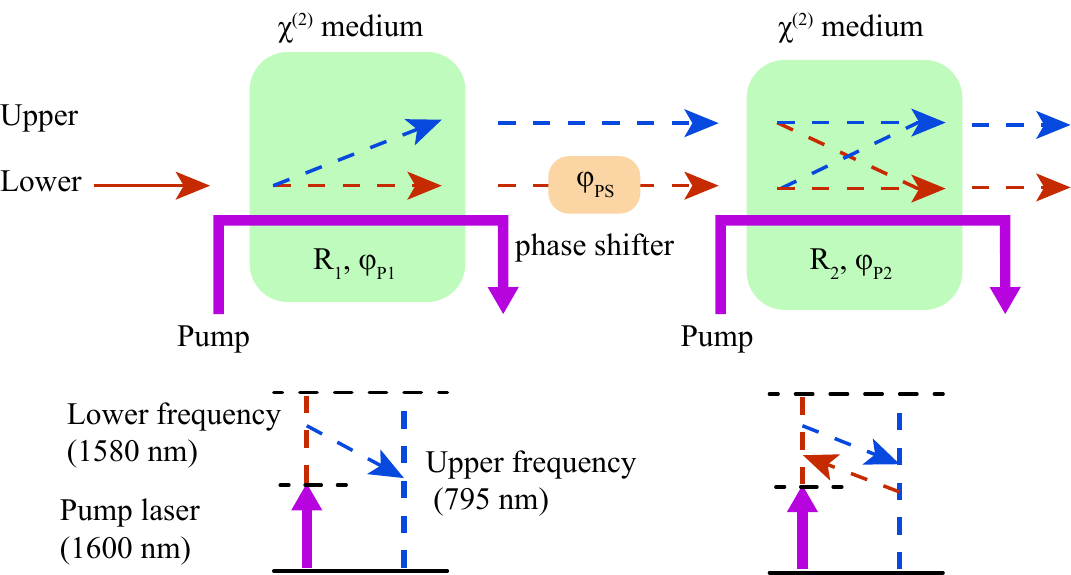}
 \end{center}
 \caption{
 The concept of frequency-domain Mach-Zehnder interferometer 
 }
 \label{fig:concept}
\end{figure*}

The frequency-domain MZI consists of two frequency-domain BSs and a phase shifter as shown in Fig.~\ref{fig:concept}.
For simplicity, we deal with a lossless optical circuit.
The phase shifter adds a phase shift $\phi_\mathrm{PS}$ on the lower frequency mode as  
$\hat{U}_\mathrm{PS}^\dagger \hat{a}_\mathrm{U} \hat{U}_\mathrm{PS} = \hat{a}_\mathrm{U}$ and 
$\hat{U}_\mathrm{PS}^\dagger \hat{a}_\mathrm{L} \hat{U}_\mathrm{PS} = e^{i\phi_\mathrm{PS}}\hat{a}_\mathrm{L}$,
where $\hat{U}_\mathrm{PS}$ is a unitary operator of the phase shifter.
The first (second) frequency-domain BS is represented by the unitary operator $U_\mathrm{BS1 (2)}$ which give the operations in Eqs.~(\ref{eq:upper}) and (\ref{eq:lower}) with reflectance $R_{1 (2)}$, transmittance $T_{1 (2)} = 1- R_{1 (2)}$ and phase  $\phi_\mathrm{P,1 (2)}$ of the pump light.
When the lower frequency light in state $\ket{\psi}_\mathrm{L}$
is injected into the interferometer,
average photon number of light 
in the upper and the lower modes is described by
\begin{eqnarray}
 p_\mathrm{U} &=& \bra{\psi}_\mathrm{L} 
  \hat{U}_\mathrm{MZI}^\dagger
 \hat{a}_\mathrm{U}^\dagger   \hat{a}_\mathrm{U}
 \hat{U}_\mathrm{MZI}
  \ket{\psi}_\mathrm{L}
=  \left\vert \sqrt{R_1 T_2}+e^{-i \delta\phi }\sqrt{T_1 R_2}\right\vert^2
\bra{\psi}_\mathrm{L}\hat{a}^\dagger_\mathrm{L} \hat{a}_\mathrm{L} \ket{\psi}_\mathrm{L}    ,\label{eq:mzi_u} \\
 p_\mathrm{L} &=& \bra{\psi}_\mathrm{L}
  \hat{U}_\mathrm{MZI}^\dagger 
 \hat{a}_\mathrm{L}^\dagger   \hat{a}_\mathrm{L}
 \hat{U}_\mathrm{MZI}
  \ket{\psi}_\mathrm{L}
   = \left\vert\sqrt{T_1 T_2} -e^{i \delta\phi}\sqrt{R_1 R_2}\right\vert^2
  \bra{\psi}_\mathrm{L}\hat{a}^\dagger_\mathrm{L} \hat{a}_\mathrm{L} \ket{\psi}_\mathrm{L},    \label{eq:mzi_l}
   \end{eqnarray}
by using Eqs.~(\ref{eq:upper}) and (\ref{eq:lower}) and the transformation of the phase shifter,
where 
$\hat{U}_\mathrm{MZI} \equiv \hat{U}_\mathrm{BS2}\hat{U}_\mathrm{PS}\hat{U}_\mathrm{BS1}$
 and $\delta \phi \equiv \phi_{\mathrm{P},1}-\phi_{\mathrm{P},2}-\phi_\mathrm{PS} $.
$p_\mathrm{U}$ and $p_\mathrm{L}$ depend on the relative phase $\delta \phi$.
This is a precise analogue to the first-order interference appearing in a conventional MZI.
We define the interference visibility by $V = (p_\mathrm{max} - p_\mathrm{min})/(p_\mathrm{max}+p_\mathrm{min})$,
where $p_\mathrm{max}$ and $p_\mathrm{min}$ are the maximum and the minimum values of $p_\mathrm{U}$ and $p_\mathrm{L}$, respectively.
From Eqs.~(\ref{eq:mzi_u}) and (\ref{eq:mzi_l}),
the visibilities of the interference fringes $V_\mathrm{U}$ and $V_\mathrm{L}$ which are observed in the upper and the lower frequency modes,
 are described by
\begin{eqnarray}
\label{eq:visU}
 V_\mathrm{U} =
   \frac{2\sqrt{R_1 T_1 R_2 T_2}}{R_1 T_2 + T_1 R_2},\\
\label{eq:visL}
 V_\mathrm{L} =
   \frac{2\sqrt{R_1 T_1 R_2 T_2}}{R_1 R_2 + T_1 T_2}.
\end{eqnarray}
The maximum visibilities $V_\mathrm{U}=V_\mathrm{L}=1$ are achieved by $R_1 = R_2 = 1/2$.

\section{Experiment}
\subsection{Experimental setup}
\begin{figure*}[t]
 \begin{center}
  \includegraphics[bb=0 0 2007 610, width=12cm]{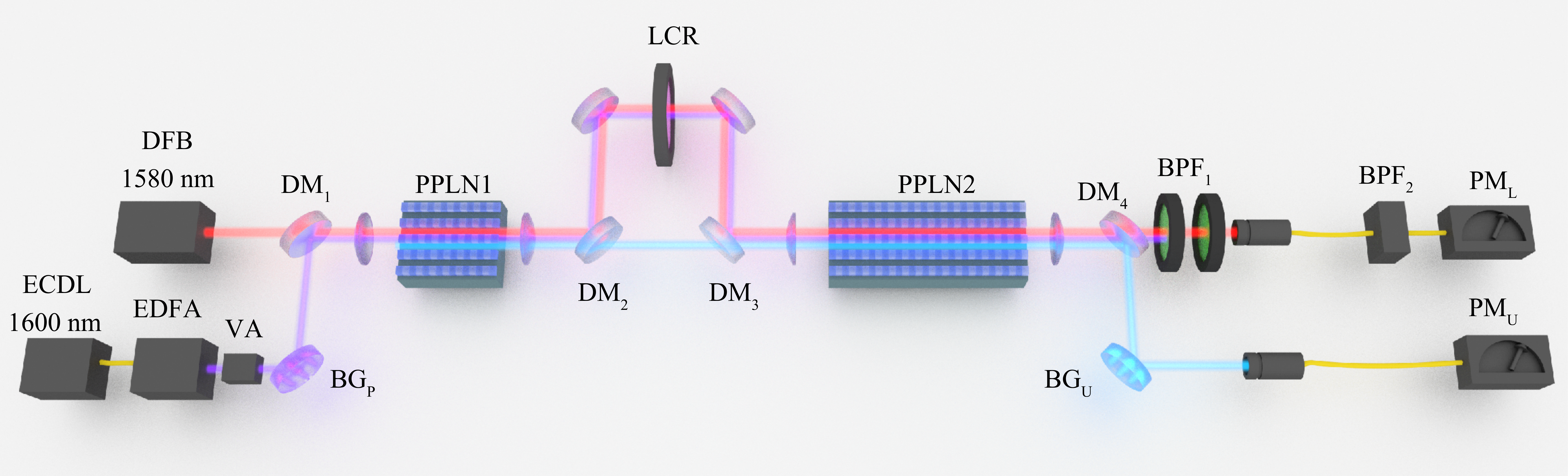}
 \end{center}
 \caption{Experimental setup. The lower frequency light~(1580~nm) is partially upconverted to the upper frequency light (795~nm)
 by three wave mixing with a strong cw pump light at 1600~nm. The upper and lower frequency lights are made to interfere at $\mathrm{PPLN_2}$ and measured by power meters $\mathrm{PM_U}$ and $\mathrm{PM_L}$.
 }
 \label{fig:setup}
\end{figure*}
The experimental setup is shown in Fig.~\ref{fig:setup}.
Two frequency-domain BSs of the MZI are composed of two PPLN waveguides \cite{Nishikawa2009} denoted by PPLN1 and PPLN2.
The PPLN1 and PPLN2 are designed for the type-0 quasi-phase-matching for three wave mixing of 1600~nm, 1580~nm and  795~nm which are used as the pump, the lower and the upper frequencies, respectively.
The pump light is  a V-polarized cw laser amplified by an Erbium-doped fiber amplifier~(EDFA).
The seed laser is an external cavity diode laser~(ECDL) with a full width at half maximum~(FWHM) of 150~kHz.
The coupled power $P_1$ of the pump beam to PPLN1 is adjusted 
by a variable attenuator~(VA) up to $\sim$~400~mW.
The pump beam is filtered by a Bragg grating~($\mathrm{BG_P}$) with a FWHM of 120~GHz
to reduce an amplified spontaneous emission noise from EDFA.
The lengths of PPLN1 and PPLN2 are 20~mm and 40~mm, respectively.
These lengths determine their FWHM acceptance widths of PPLN1 and PPLN2 
 as 140~GHz and 70~GHz, respectively,
for the conversion between 795~nm and 1580~nm.

The input of the present MZI interferometer is a V-polarized lower frequency light (1580~nm) taken from a  cw distributed feedback~(DFB) laser with the linewidth of 10 MHz~(FWHM) and the power of 10~mW.
The input light is combined with the pump light by a dichroic mirror~($\mathrm{DM_1}$)
and the two lights are coupled to PPLN1.
The upper frequency light~(795~nm) generated at PPLN1 is separated from
 the lower frequency and pump lights by $\mathrm{DM_2}$.
The LCR adds approximately the same phase shifts to the lower-frequency and pump lights
to sweep the relative phase $\delta \phi$ in Eqs.~(\ref{eq:mzi_u}) and (\ref{eq:mzi_l}).
Then, three lights are recombined by $\mathrm{DM_3}$,
and coupled to PPLN2.

Among the outputs from PPLN2, the upper frequency light (795~nm) is picked up by $\mathrm{DM_4}$ and $\mathrm{BG_U}$ with the FWHM bandwidth of $\Delta_\mathrm{f,U} \equiv 99$~GHz.
The lower frequency light~(1580~nm) is separated from the pump beam at 1600~nm by bandpass filters~($\mathrm{BPF_1}$,  $\mathrm{BPF_2}$).
The total bandwidth of the set of two bandpass filters is $\Delta_\mathrm{f,L} = 69$~GHz.
The upper and lower frequency lights extracted  are coupled to single mode fibers and measured by power meters $\mathrm{PM_U}$ and  $\mathrm{PM_L}$.

\subsection{Experimental results}
\begin{figure}[t]
 \begin{center}
  \includegraphics[bb=0 0 1096 361, width=12cm]{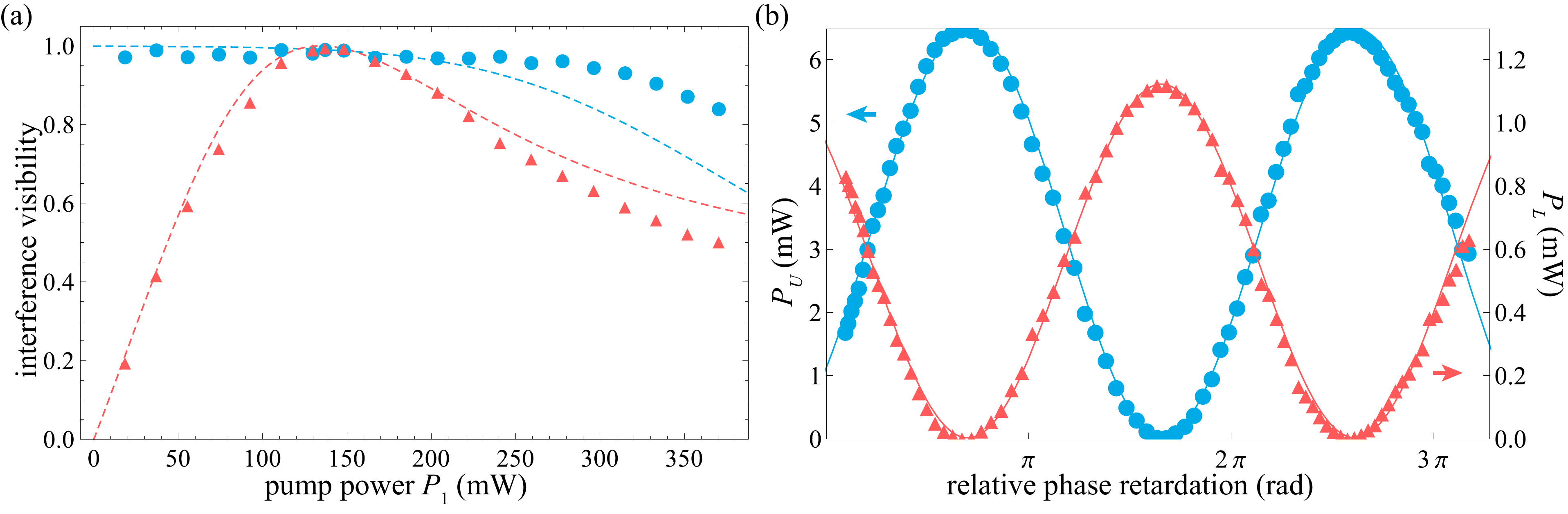}
 \end{center}
 \caption{ (a) 
 Pump power dependencies of the interference visibility of the upper (circles) and the lower (triangles) frequency modes.
 (b) The observed interference fringes of upper (circles) and lower (triangles) frequency modes at $P_1=140$~mW.
 }
 \label{fig:fringe}
\end{figure}
We measured the dependencies of the interference visibility of 
the upper mode $V_\mathrm{U}(P_1)$ and the lower mode $V_\mathrm{L}(P_1)$ on the pump power $P_1$
from the output power of the 795-nm and 1580-nm light by varying the relative phase through LCR.
The result is shown in Fig.~\ref{fig:fringe}~(a).
The observed interference fringes at $P_1=140$~mW are shown in Fig.~\ref{fig:fringe}~(b).
The interference fringes in the upper and the lower frequency modes are clearly seen.
The observed interference visibilities take the same value of 0.99 in both modes at the pump power of 140~mW.

\subsection{Theoretical analysis}
\begin{figure*}[t]
 \begin{center}
  \includegraphics[bb=0 0 350 144]{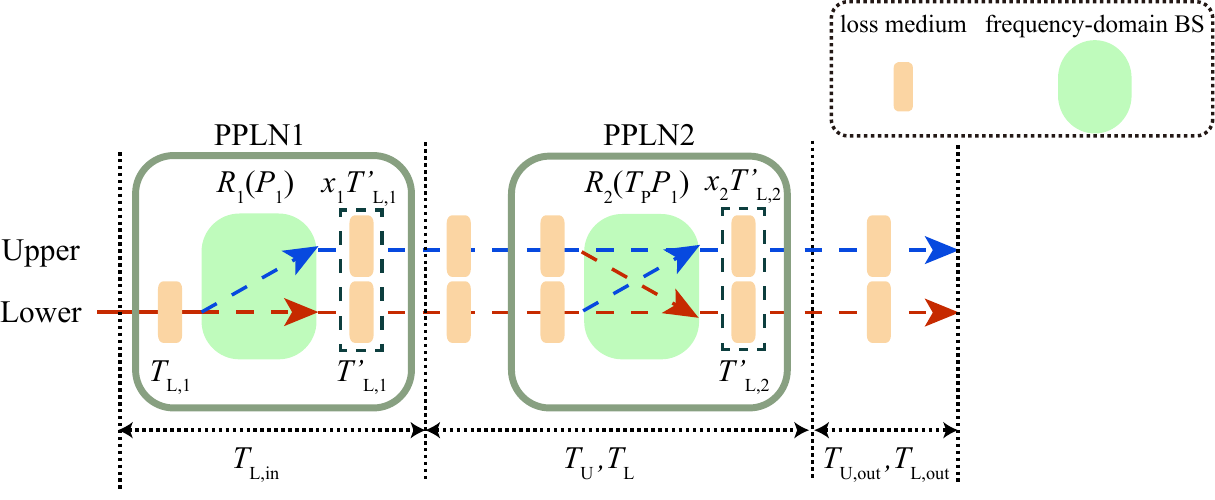}
 \end{center}
 \caption{
 The theoretical model with lossless frequency-domain BSs and loss media which represent the observed transmittance and virtual loss.
}
  \label{fig:model}
\end{figure*}
The visibilities $V_\mathrm{U}(P_1)$ and $V_\mathrm{L}(P_1)$ depend not only on the pump power $P_1$ but also on the losses inside the frequency-domain MZI.
In Fig.~\ref{fig:model}, we show a theoretical model of a lossy frequency-domain MZI.
In order to characterize the losses, we consider virtual loss media in front of and behind lossless frequency-domain BSs with conversion efficiencies $R_1(P_1)=1-T_1(P_1)$ and $R_2(T_\mathrm{P}P_1)=1-T_2(T_\mathrm{P}P_1)$.
The ratios of transmittances of virtual loss media behind frequency-domain BSs in upper mode to that in lower mode are 
denoted by $x_1$ for PPLN1 and $x_2$ for PPLN2.
The transmittances of PPLN1 for the 1580-nm light, when there is no pump light, is denoted by $T_\mathrm{L,in}=T_\mathrm{L,1}T'_\mathrm{L,1}$, 
where $T_\mathrm{L,1}$ and $T'_\mathrm{L,1}$ represent the transmittance of loss media 
in front of and behind lossless frequency-domain BSs.
The transmittances from the end of PPLN1 to the end of unpumped PPLN2
are denoted by $T_\mathrm{U}$ and $T_\mathrm{L}$ for the 795-nm and 1580-nm light, respectively.
For the pump light, we ignore the depletion resulting from the wavelength conversion and assume that the pump power for the PPLN2 to that for PPLN1 is given by a constant factor $T_\mathrm{P}$.
We estimated the values of $T_\mathrm{L,in}$, $T_\mathrm{U}$, $T_\mathrm{L}$, and $T_\mathrm{P}$
directly in our experiment. 
The result are $T_\mathrm{L,in}\approx 0.64$, $T_\mathrm{U}\approx 0.63$, $T_\mathrm{L}\approx 0.60$, $T_\mathrm{P}\approx 0.54$.
The visibility of the lossy MZI for the output of the upper frequency mode is
obtained by replacing $T_1$, $R_1$, $T_2$ and $R_2$ with $T_1(P_1)T_\mathrm{L,in}$, $R_1(P_1)T_\mathrm{L,in}x_1$, $T_2(T_\mathrm{P}P_1)T_\mathrm{U}$ and $R_2(T_\mathrm{P}P_1)T_\mathrm{L} x_2$ in Eq.~(\ref{eq:visU}), respectively, 
as
\begin{eqnarray}
\label{eq:visUtrans}
 V_\mathrm{U} (P_1)&=&
   \frac{2\sqrt{R_1(P_1) T_1(P_1) R_2(T_\mathrm{P} P_1) T_2(T_\mathrm{P} P_1)  T_\mathrm{U} T_\mathrm{L}x_1x_2 }}{R_1(P_1) T_2(T_\mathrm{P} P_1)T_\mathrm{U} x_1 + T_1(P_1)  R_2(T_\mathrm{P} P_1) T_\mathrm{L}x_2 },
\end{eqnarray}
In the same manner,
a visibility for the lower frequency mode is obtained by replacing 
$T_1$, $R_1$, $T_2$ and $R_2$ with $T_1(P_1)T_\mathrm{L,in}$, $R_1(P_1)T_\mathrm{L,in}x_1$, $T_2(T_\mathrm{P}P_1)T_\mathrm{L}$ and $R_2(T_\mathrm{P}P_1)T_\mathrm{U}/ x_2$ in Eq.~(\ref{eq:visL}), respectively, 
as
\begin{eqnarray}
\label{eq:visLtrans}
  V_\mathrm{L} (P_1)&=&
   \frac{2\sqrt{R_1(P_1) T_1(P_1) R_2(T_\mathrm{P} P_1) T_2(T_\mathrm{P} P_1) T_\mathrm{U}  T_\mathrm{L} x_1 /x_2 }}{R_1(P_1)  R_2(T_\mathrm{P} P_1) T_\mathrm{U}x_1 /x_2 + T_1(P_1)  T_2(T_\mathrm{P} P_1) T_\mathrm{L}}.
\end{eqnarray}

\begin{figure}[t]
 \begin{center}
  \includegraphics[bb=0 0 974 341,  width=12cm]{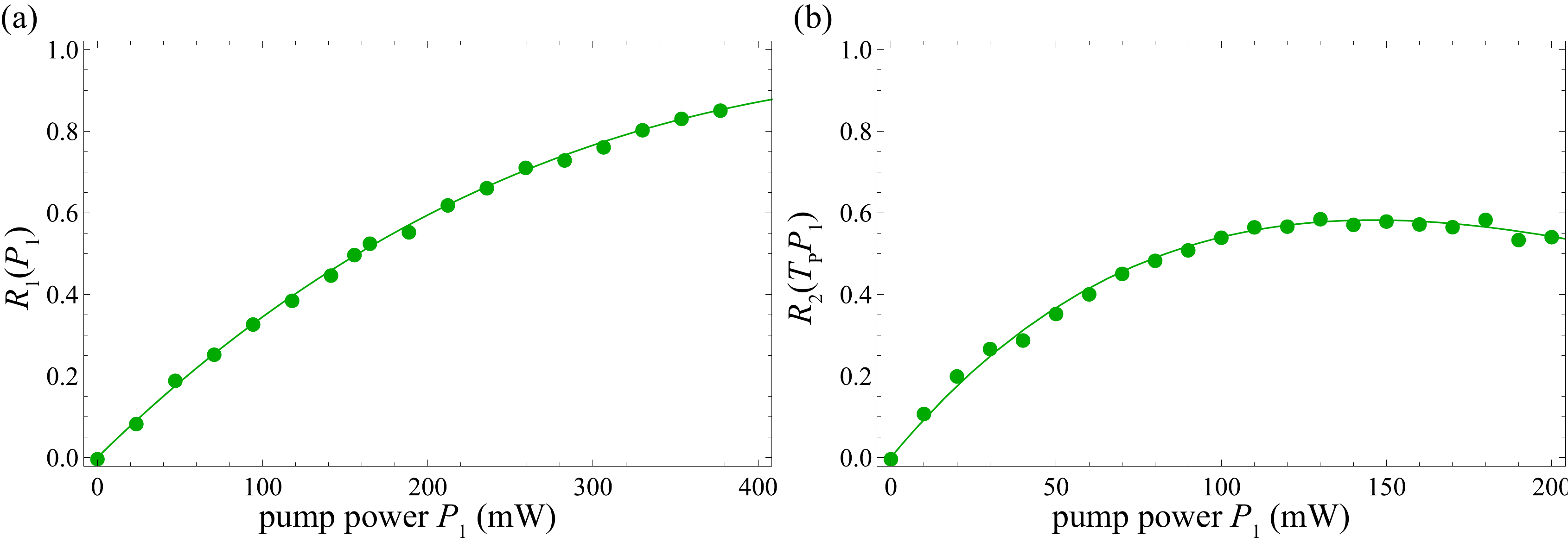}
 \end{center}
 \caption{ Internal conversion efficiencies of (a) PPLN1 and (b) PPLN2. 
 The curves are obtained by the best fit to $R_1(P_1)$ and $R_2(T_\mathrm{P} P_1)$ with $A \sin^2(\sqrt{\eta P})$,
  where the fitting parameters $A$ and $\eta$ have been estimated to be 0.94 and 0.0042 / mW for $R_1(P_1)$
  and 0.58 and 0.017 / mW for $R_2(T_\mathrm{P} P_1)$.
    }
 \label{fig:conv}
\end{figure}
In order to calculate $V_\mathrm{U} (P_1)$ and $V_\mathrm{L} (P_1)$, 
we estimated the internal conversion efficiency $R_1(P_1)=1-T_1(P_1)$ and the parameter $x_1$ of PPLN1 
in the same manner as in Ref.~\cite{Ikuta2013:3}, by observing 
the output powers of 795-nm and 1580-nm light by power meters
$\mathrm{PM_U}$ and $\mathrm{PM_L}$ while varying the pump power $P_1$. 
The temperature of  PPLN2 was chosen to make its conversion efficiency negligible.
Taking into account the values of $T_\mathrm{L}$ and $T_\mathrm{U}$ above, 
we determined $x_1=1.4\pm 0.1$ and $R_1(P_1)$ as shown in Fig.~\ref{fig:conv}(a).
The best fit to $R_1(P_1)$ with $A \sin^2(\sqrt{\eta P_1})$ gives $A\approx 0.94$ and $\eta \approx 0.0042$~/mW.
From the result, $R_1(P_1)\sim 0.5$ is achieved at 140-mW pump power.
In a similar way, we determined $x_2 = 1.0 \pm 0.1$ and $R_2(T_\mathrm{P} P_1)$ versus the
pump power $P_1$ as shown in Fig.~\ref{fig:conv}(b). 
The best fit to $R_2(P)$ with $A \sin^2(\sqrt{\eta P})$ gives $A\approx 0.58$ and $\eta \approx 0.017$~/mW.

By using observed values,
we calculated  $V_\mathrm{U} (P_1)$ and $V_\mathrm{L} (P_1)$ in Eqs.~(\ref{eq:visUtrans}) and (\ref{eq:visLtrans}) as shown with dashed curves in Fig.~\ref{fig:fringe}~(a).
These are in good agreement with the experimental data around the pump power of 140~mW.
In our lossy interferometer, the transmittances $T_\mathrm{U}$ and $T_\mathrm{L}$ for the upper and lower frequency modes have similar values, and thus the maximum visibilities of $V_\mathrm{U}$ and $V_\mathrm{L}$ are 
simultaneously achieved at 140-mW pump power corresponding to
$R_1(P_1)\sim 0.5$ and $R_2(T_\mathrm{P} P_1)\sim 0.5$,
similarly to the lossless case.

\section{Discussion}
\begin{figure*}[t]
 \begin{center}
  \includegraphics[bb=0 0 973 318, width=12cm]{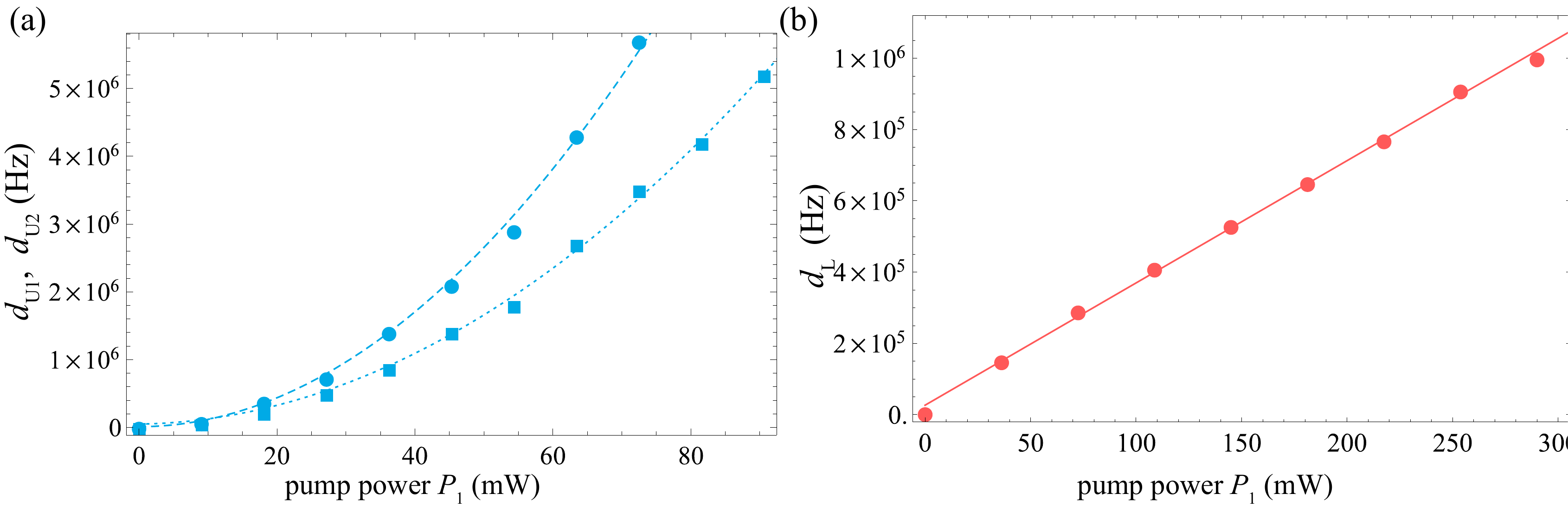}
 \end{center}
 \caption{ Dependencies of the background noise photons on the pump power $P_1$. 
 (a) The circles and squares represent the background noises $d_\mathrm{U,1}$ and $d_\mathrm{U,2}$ at 795~nm generated from PPLN1 and PPLN2, respectively.
 The dashed and dotted curves is obtained by the best fit to the experimental data with $A P_1^2 + B P_1 + C$,
where the fitting parameters $A$, $B$ and $C$ are
 $3.5\times 10^3$ / mW${}^2$, $4.6 \times 10^2$ / mW  and $9.1 \times 10^3$ for $d_\mathrm{U,1}$, 
 and are
 $2.0\times 10^3 / \mathrm{mW}^2$,  $2.9 \times 10^3 $ / mW and $4.8 \times 10^4$ for $d_\mathrm{U,2}$.
 (b) The circles represent the background noises $d_\mathrm{L}$ at 1580~nm.
 The solid curve is obtained by the best fit to the experimental data with $A P_1 + B$,
  where the fitting parameters $A$ and $B$ are $6.2\times 10^3$ per mW and $2.6\times 10^4$.
}
  \label{fig:noise}
\end{figure*}
\begin{figure*}[t]
 \begin{center}
  \includegraphics[bb=0 0 1080 391, width=12cm]{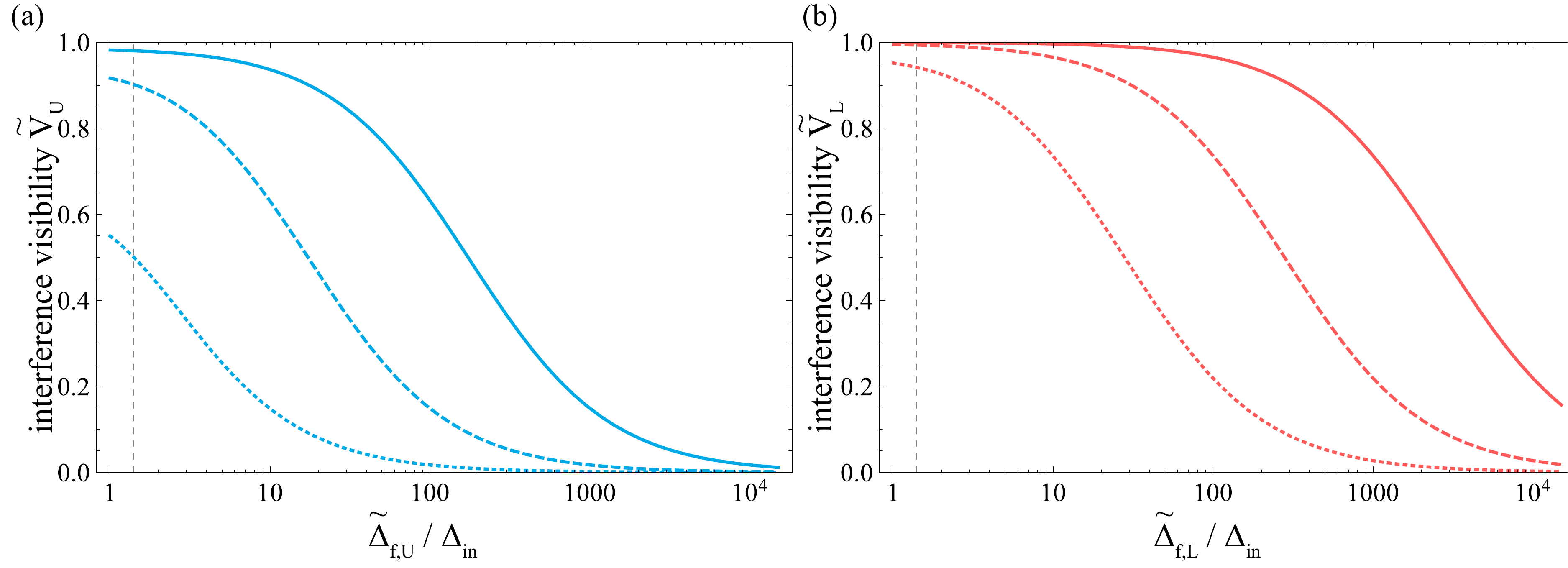}
 \end{center}	
 \caption{The expected interference visibilities in (a) the upper and (b) the lower frequency modes for a single photon as a function of $\widetilde{\Delta}_\mathrm{f,U}/\Delta_\mathrm{in}$ and $\widetilde{\Delta}_\mathrm{f,L}/\Delta_\mathrm{in}$ in the cases of the average photon number of 1 (solid curve), 0.1 (dashed curve) and 0.01 (dotted curve) at $P=140$~mW. 
 The vertical dashed lines represents the bandwidth of $1.4 \times \Delta_\mathrm{in}$.
    }
 \label{fig:vis}
\end{figure*}
In this section, we discuss the performance of the frequency-domain MZI expected 
in the case where the input comes from a typical single photon source.
We first measured background photon count rates by replacing the power meters with superconducting single-photon detectors~(SSPDs)\cite{Miki2013}.
The quantum efficiencies are 0.3 and 0.6 for the 795-nm and 1580-nm light, respectively.
For measuring the background photon count rate $d_\mathrm{U1}$ at 795 nm induced by PPLN1, we dumped the 1580-nm and 1600-nm light after PPLN1 and injected only the 795-nm light to PPLN2.
For measuring the background photon count rate $d_\mathrm{U2}$ at 795 nm induced by PPLN2, we dumped the 795-nm light after PPLN1 and injected the 1580-nm and 1600-nm light to PPLN2.
For background photons at 1580 nm, we cannot separately measure the noise photons induced by pump light at PPLN1 and PPLN2.
Therefore we measured only the sum of the background photon count rates $d_\mathrm{L}$ induced by pump light at PPLN1 and PPLN2.
The experimental result is shown in Fig.~\ref{fig:noise}.
The result shows nonlinear and linear dependencies for 795-nm and 1580-nm light, which is the same tendency reported in Ref.~\cite{Ikuta2013:3}.
It means the background photons are mainly originated from Raman scattering of the strong pump light.

By using the observed background photon count rates,
we estimate the expected interference visibility when we use  typical single photon source and filters.
We assume that the spectral bandwidth of the input light is a transform-limited pulse of a Gaussian 
with its FWHM bandwidth of $\Delta_\mathrm{in}$ and the measurement time window
 is set to 
 $4 \ln 2  /\pi \Delta_\mathrm{in} $ which corresponds to
the twice of FWHM pulse length of the input light, which leads to 98\% collection efficiency.
The background-photon reduction of the output light is performed by the filters
with the bandwidth of $\widetilde{\Delta}_\mathrm{f,U}$ and $\widetilde{\Delta}_\mathrm{f,L}$ for the upper light and the lower light, where $\widetilde{\Delta}_\mathrm{f,U}, \widetilde{\Delta}_\mathrm{f,L} > \Delta_\mathrm{in}.$
We assume that the spectral shapes of the filters are Gaussian with the peak value of unity and the transmittances $\widetilde{T}_\mathrm{U,out}$ and $\widetilde{T}_\mathrm{L,out}$ from PPLN2 to the single mode fibers 
are estimated by using $\Delta_\mathrm{in},  \widetilde{\Delta}_\mathrm{f,U}$, $\widetilde{\Delta}_\mathrm{f,L}$ and the observed values $T_\mathrm{U,out}\approx 0.71$ and $T_\mathrm{L,out}\approx 0.13$.
It is reasonable to assume the numbers of the background photons are proportional to 
the measurement time window 
and the bandwidth.
Under this assumption, the numbers of the background photons in the measurement time window
are derived  as functions of $P_1$, $\widetilde{\Delta}_\mathrm{f,U}/\Delta_\mathrm{in}$ and $\widetilde{\Delta}_\mathrm{f,L}/\Delta_\mathrm{in}$, as follows; 
\begin{eqnarray}
 n_\mathrm{U}(P_1, \widetilde{\Delta}_\mathrm{f,U}/\Delta_\mathrm{in}) &=& (d_\mathrm{U,1}(P_1)+d_\mathrm{U,2}(P_1)) \frac{ 4 \ln 2}{\pi \Delta_\mathrm{f,U} } \frac{\widetilde{\Delta}_\mathrm{f,U}}{\Delta_\mathrm{in}},\\
 n_\mathrm{L}(P_1, \widetilde{\Delta}_\mathrm{f,L}/\Delta_\mathrm{in}) &=&
 d_\mathrm{L}(P_1) \frac{4 \ln 2}{\pi \Delta_\mathrm{f,L}}\frac{\widetilde{\Delta}_\mathrm{f,L}}{\Delta_\mathrm{in}}.
\end{eqnarray}
By using the experimental parameters including the conversion efficiencies, the transmittances of the optical circuit and the quantum efficiencies of SSPDs, 
the expected visibilities with the average photon number $\mu$ of the input pulse are represented by
\begin{eqnarray}
\label{eq:visUbg}
 \widetilde{V}_\mathrm{U} (P_1, \widetilde{\Delta}_\mathrm{f,U}/\Delta_\mathrm{in})=
   \frac{2\sqrt{R_1(P_1) T_1(P_1) R_2(T_\mathrm{P} P_1) T_2(T_\mathrm{P} P_1) T_\mathrm{U} T_\mathrm{L}x_1x_2 } }
   {R_1(P_1)T_2(T_\mathrm{P} P_1) T_\mathrm{U}x_1  + T_1(P_1)  R_2(T_\mathrm{P} P_1)T_\mathrm{L} x_2 +n_\mathrm{U}(P_1, \widetilde{\Delta}_\mathrm{f,U}/\Delta_\mathrm{in})/(\mu T_\mathrm{L,in} \widetilde{T}_\mathrm{U,out}) },\\
\label{eq:visLbg}
  \widetilde{V}_\mathrm{L} (P_1,\widetilde{\Delta}_\mathrm{f,L}/\Delta_\mathrm{in})=
   \frac{2\sqrt{R_1(P_1) T_1(P_1) R_2(T_\mathrm{P} P_1) T_2(T_\mathrm{P} P_1) T_\mathrm{U} T_\mathrm{L} x_1/x_2 }}
{R_1(P_1) R_2(T_\mathrm{P} P_1)T_\mathrm{U} x_1/x_2 +T_1(P_1) T_2(T_\mathrm{P} P_1)T_\mathrm{L}+n_\mathrm{L}(P_1, \widetilde{\Delta}_\mathrm{f,L}/\Delta_\mathrm{in} )/(\mu T_\mathrm{L,in}\widetilde{T}_\mathrm{L,out})}.
\end{eqnarray}
The estimated visibilities at $P_1=140$~mW with $\mu=1, 0.1, 0.01$ are shown in Fig.~\ref{fig:vis} by using $T_\mathrm{L,in} \approx 0.64$.
We see that the bandpass filters with the bandwidth of $1.4 \times \Delta_\mathrm{in}$ is sufficient for achieving high visibilities over 0.9 when the input light has the average photon number above 0.1. 
For example, in the case of a single photon with a bandwidth of 9.6~MHz borrowed from Ref.~\cite{Bao2008},
a visibility of $>0.9$ will be obtained by using filters with a bandwidth of $<13$~MHz.
If we use a single photon with a bandwidth of 1.2~GHz reported in Ref.~\cite{Halder2008},
the high visibility will be obtained by using filters with a bandwidth of $<1.6$~GHz.
Such frequency filters are achieved by commonly-used etalons, cavities and fiber Bragg gratings.

\section{Conclusion}
In conclusion, we have demonstrated the first-order interference between 1580-nm and 795-nm light by constructing the frequency-domain MZI composed of two partial frequency converters using PPLN waveguide and the pump light at 1600~nm  as the frequency-domain BSs.
The observed visibilities for both output modes~(1580~nm and 795~nm) are 0.99
when the conversion efficiency is set to be $\sim$ 0.5.
The background noise estimation of our interferometer shows that if the transform-limited light pulse with the average photon number over 0.1 and sufficiently narrow filters are used, the visibility will be still over 0.9.
Such condition can be realized by current single photon sources and usual filtering techniques.
We believe that the interferometer will be useful tool for large-scale frequency-encoded photonic quantum information processing.

\section*{Acknowledgments}
Core Research for Evolutional Science and
Technology, Japan Science and Technology Agency (CREST,
JST); Ministry of Education, Culture, Sports, Science, and
Technology (MEXT); Japan Society for the Promotion of
Science (JSPS) (JP25247068, JP15H03704,
JP16H02214, JP16K17772); JSPS Grant-in-Aid for JSPS
Fellows (JP14J04677).

\end{document}